\documentclass{article}
\usepackage{amssymb}
\usepackage{amsmath}

\begin{document}


\title{The Mass and Velocity of Light from Energy and Momentum Conservation}

\author{Volodimir Simulik $^{1}$, Denys I. Bondar $^{2}$\\
$^{1}$Institute of Electron Physics of the NAS of Ukraine\\
21 Universitetska Str. Uzhgorod 88017, Ukraine\\
Email: vsimulik@gmail.com\\
$^{2}$Department of Physics and Engineering Physics\\
Tulane University, New Orleans, Louisiana 70118, United States\\
Email: dbondar@tulane.edu}


\maketitle

\begin{abstract}
While the numerical value of the speed of light is known with extraordinary precision, its theoretical definition remains a subject of fundamental interest. We show that the definition of mass and velocity of light follow from the conserved quantities of the electromagnetic field. The proposed definition of the speed of light is always bounded from above by the phase velocity and equals it for plane waves. As a consequence, we obtain a generalization of Einstein's mass-energy relation for electromagnetic fields in media: $m = \varepsilon\mu E / c^2$. Hence, irrespective of the light's intensity, the electromagentic field in near-zero-index material is always massless. This approach offers new pedagogical insights into the fundamental nature of light propagation.
\end{abstract}


\pagestyle{plain}
\pagenumbering{arabic}
\setcounter{page}{1}

\section{Introduction}

The velocity of light stands as one of the most fundamental constants in physics, yet its theoretical definition continues to evolve beyond the well-established experimental determinations. From the early theoretical considerations of Gauss (1813), Ampère (1827), Faraday (1831), Lenz (1834), Maxwell (1865), and Einstein (1905), the quest for a rigorous mathematical definition for the speed of light has engaged physicists for over two centuries.

The experimental determination of light's speed spans centuries of increasingly precise measurements, from Rømer (1676) and Bradley (1726) through Fizeau (1849), Michelson (1926), and culminating in the landmark precision measurements by Evenson et al. (1972–1973) \cite{Real-1,Real-2,Real-3,Real-4}. This rich experimental history has been comprehensively reviewed in \cite{Real-5}. Despite numerous attempts since 1973 to refine our understanding of this fundamental constant \cite{Real-6,Real-7}, most approaches focus on measurement techniques rather than theoretical foundations.

There are at least eight definitions of the velocity of light
\cite{Real-18, Real-21}. Traditional definitions of light velocity include phase velocity, group velocity, and energy transport velocity \cite{Real-8,Real-9,Real-10,Real-11,Real-12,Real-19}. For a monochromatic wave characterized by $\Psi(t,\vec{x}) = A(\vec{x})\cos(\omega t - g(\vec{x}))$, the phase velocity is $v_{\text{phase}} = \omega/|\nabla g|$, which reduces to $v_{\text{phase}} = \omega/k$ for plane waves. The group velocity is $v_{\text{group}} = d\omega/dk$. However, these definitions do not specify which equation $\Psi$ satisfies. The energy transport velocity~\cite{Real-20} is $\vec{v}_{\text{energy}} = \vec{S}/U$, where $\vec{S}$ is the Poynting vector and $U$ is the energy density. But, as noted in \cite{Real-18}, the velocity of energy transport presents experimental measurement challenges.

For broader context, Adler provides a historical and pedagogical discussion of the notion of mass in relativity \cite{Real-22}, while Mendelson gives an insightful history of the notation $c$ \cite{Real-23}. 

In this paper, we propose a novel approach: defining the mass and velocity of light as integral characteristics of the electromagnetic field derived directly from conservation laws. Our main result is that these fundamental properties follow naturally from the conserved quantities of energy and momentum. This formulation expresses light velocity as a functional of the electric and magnetic field strengths, providing a direct connection between the dynamical properties of electromagnetic fields and their propagation characteristics.

The paper is organized as follows. Section~\ref{SecDimensionalAnalysis} infers the energy-mass equivalence from dimensional analysis of Maxwell's conservation laws. Section~\ref{SecScwhingerDeriv} provides a rigorous derivation of the mass-energy equivalence and velocity of light in vacuum following Schwinger's approach. Section~\ref{SecMedium} extends these concepts to non-conducting media. The final section draws conclusions and discusses implications for modern optics, particularly near-zero-index materials.

\section{Inferring $E=mc^2$ from dimensional analysis of Maxwell's equations}\label{SecDimensionalAnalysis}

Recall that free Maxwell's equations in vacuum read
\begin{align}\label{EqMaxwellMicro}
    & \nabla\times \vec{B}^{vac}  = \frac{1}{c}\frac{\partial}{\partial t} \vec{E}^{vac}, \notag\\
    & \nabla\times \vec{E}^{vac}  = -\frac{1}{c}\frac{\partial}{\partial t} \vec{B}^{vac}, \notag\\
    & \nabla\cdot \vec{B}^{vac} = \nabla\cdot \vec{E}^{vac} = 0.
\end{align}
At this point, the constant $c$ in Maxwell's equations is merely the ratio of electrical units, as described in Sec.~II of Ref.~\cite{Real-23}. We will derive in the following section that $c$ also equals the speed of light in vacuum.

As shown in textbooks on electrodynamics, these equations obey the conservation of energy in integral and differential forms
\begin{align}
    -\frac{d}{dt} \int_V U^{vac} d^3 x = \oint_{\partial V} \vec{S}^{vac} \cdot d\vec{s}, \label{EqEnergyConservation}\\
    \frac{\partial U^{vac}}{\partial t} + \nabla \cdot \vec{S}^{vac} = 0, \label{EqDiffEnergyConservation}
\end{align}
where $U^{vac}$ denotes the energy density and $\vec{S}^{vac}$ denotes the Poynting vector in vacuum,
\begin{align}
    U^{vac} &= \left( \vec{E}^{vac} \cdot \vec{E}^{vac} + \vec{B}^{vac} \cdot \vec{B}^{vac} \right) / (8\pi), \label{EqDefUvca} \\
    \vec{S}^{vac} &= c \vec{E}^{vac} \times \vec{B}^{vac} / (4\pi),
\end{align}
as well as the conservation of momentum (in differential form)
\begin{align}
    \frac{\partial}{\partial t} G^{vac}_k +  \sum_{l=1}^3 \frac{\partial}{\partial x_l} T^{vac}_{kl} = 0, \label{EqMomentumConservation}\\
    \vec{S}^{vac} = c^2 \vec{G}^{vac},  \label{EqSGc2}
\end{align}
where $\vec{G}^{vac}$ is the momentum density vector and $T^{vac}_{kl}$ is the stress tensor.

The unit of $\vec{S}^{vac}$ is energy flux, i.e.,
\begin{align}
    [\vec{S}^{vac}] = \frac{\text{(energy)}}{\text{(area)} \cdot \text{(time)}}
\end{align}
and
\begin{align}
    [\vec{G}^{vac}] = \frac{\text{(momentum)}}{\text{(volume)}}.
\end{align}
With this in mind, as far as dimensions are concerned, Eq.~\eqref{EqSGc2} implies
\begin{align}
    \frac{\text{(energy)}}{\text{(area)} \cdot \text{(time)}} &= \frac{\text{(momentum)}}{\text{(volume)}} c^2 \Longrightarrow  \\
    \frac{\text{(energy)}}{\text{(area)} \cdot \text{(length)}} \cdot \frac{\text{(length)}}{\text{(time)}} &= \frac{\text{(energy)}}{\text{(volume)}} \cdot \text{(velocity)} \\
    & =\frac{\text{(mass)} \cdot \text{(velocity)}}{\text{(volume)}} c^2 \Longrightarrow  \\
    \text{(energy)} &= \text{(mass)} c^2.
\end{align}
The last equation suggests the mass-energy equivalence for the electromagnetic field. We can turn this into the definition of the mass of light in vacuum
\begin{align}\label{EqMassVacuum}
    m^{vac} = E^{vac} / c^2,
\end{align}
where $E^{vac}$ is the total energy of the electromagnetic field obtained from the energy density
\begin{align}\label{EqDefEvac}
    E^{vac}  =  \int U^{vac} d^3 x.
\end{align}
For the history and ideas about the electromagnetic nature of mass, see Chapter 11 of monograph~\cite{Real-24}.

\section{Schwinger's derivation of $E=mc^2$ from Maxwell's equations in vacuum}\label{SecScwhingerDeriv}

We emphasize that Eq.~\eqref{EqMassVacuum} was inferred rather than rigorously derived. We now reproduce the mathematically rigorous derivation put forth by Schwinger in Sec.~3.4 of his textbook~\cite{Real-25}.

The only assumption required for this derivation is that the electromagnetic field vanishes at infinity. This is readily accomplished if the wave packet is initially localized in space.

For $k \in \{1,2,3\}$, calculate $x_k \cdot$ Eq.~\eqref{EqDiffEnergyConservation} $ - c^2 t \cdot$ Eq.~\eqref{EqMomentumConservation} to get
\begin{align}\label{EqSchwinger1}
    x_k \left( \frac{\partial U^{vac}}{\partial t} + \nabla \cdot \vec{S}^{vac} \right)
    - c^2 t \left( \frac{\partial}{\partial t} G^{vac}_k +  \sum_{l=1}^3 \frac{\partial}{\partial x_l} T^{vac}_{kl} \right) = 0.
\end{align}
Using the product rule and the fact that $\partial x_k / \partial x_l = \delta_{k, l}$, we can show that the following equation is equivalent to Eq.~\eqref{EqSchwinger1}:
\begin{align}
    \frac{\partial}{\partial t}\left( x_k U^{vac} - c^2 t G_k^{vac} \right) + \sum_{l=1}^3 \frac{\partial}{\partial x_l} \left( x_k S_l^{vac} - c^2 t T_{kl}^{vac} \right) = 0.
\end{align}
Next, we integrate this equation over all space. Since the second term is in the form of a divergence, according to the divergence theorem, it vanishes as the fields equal zero at the boundary.
\begin{align}
    & \int d^3 x \frac{\partial}{\partial t}\left( x_k U^{vac} - c^2 t G_k^{vac} \right) = 0 \Longrightarrow  \\
    & \frac{d}{dt} \int d^3 x \left( \vec{x} U^{vac} - c^2 t \vec{G}^{vac} \right) = \vec{0}. \label{EqSchwinger2}
\end{align}

Since the energy density [Eq.~\eqref{EqDefUvca}] is non-negative and the total energy of the field~\eqref{EqDefEvac} is positive, the quantity
\begin{align}
    \mathcal{P} = U^{vac} / E^{vac}
\end{align}
obeys the properties
\begin{align}
    \int \mathcal{P} d^3 x = 1, \qquad 0 \leq \mathcal{P} \leq 1.
\end{align}
Hence, $\mathcal{P}$ can be interpreted as a probability distribution, with which we can introduce the mean position of the electromagnetic wave packet
\begin{align}
     \langle \vec{x} \rangle^{vac} = \int \vec{x}  \mathcal{P} d^3x.
\end{align}
Therefore, Eq.~\eqref{EqSchwinger2} implies
\begin{align}
    \frac{d}{dt} \left( E^{vac} \langle \vec{x} \rangle^{vac} - c^2 t \vec{P}^{vac} \right) = \vec{0},
\end{align}
where $\vec{P}^{vac}$ denotes the total momentum of the field
\begin{align}\label{EqSchwinger3}
    \vec{P}^{vac} =  \int \vec{G}^{vac} d^3 x = \int \frac{\vec{S}^{vac}}{c^2} d^3 x.
\end{align}
Both $E^{vac} = \text{const}$ and $\vec{P}^{vac} = \text{const}$ are conserved quantities according to Eqs.~\eqref{EqEnergyConservation} and~\eqref{EqMomentumConservation}. This yields
\begin{align}\label{EqFirstEhrenfestVacuumTh}
    & \frac{E^{vac}}{c^2}  \frac{d}{dt} \langle \vec{x} \rangle^{vac} = \vec{P}^{vac},
\end{align}

Equation~\eqref{EqFirstEhrenfestVacuumTh} closely resembles the celebrated Ehrenfest theorem~\cite{Real-26, Real-26b, Real-26c}, $m d\langle \hat{x} \rangle / dt = \langle \hat{p} \rangle$, which connects the expectation values of position and momentum for a quantum particle through mass. Using this analogy, Eq.~\eqref{EqFirstEhrenfestVacuumTh} implies the mass-energy equivalence~\eqref{EqMassVacuum}.

Also, Eq.~\eqref{EqFirstEhrenfestVacuumTh} can be interpreted as a definition of the speed of light,
\begin{align}\label{EqSchwingerDefSpeedOfLight}
    \frac{d}{dt} \langle \vec{x} \rangle^{vac} = c^2 \frac{\vec{P}^{vac}}{E^{vac}}.
\end{align}
It follows from Eq.~\eqref{EqVelocityInMediumEqualsPhaseVel}, derived below, that
\begin{align}
    \left| \frac{d}{dt} \langle \vec{x}  \rangle^{vac} \right| = c.
\end{align}

\section{The energy-mass equivalence and velocity of light in medium}\label{SecMedium}

Macroscopic Maxwell's equations in medium with no free charges and currents read
\begin{align}\label{EqMaxwellMacro}
    & \nabla\times \vec{H} = \frac{1}{c} \frac{\partial}{\partial t} \vec{D}, \notag\\
    & \nabla\times \vec{E} = -\frac{1}{c} \frac{\partial}{\partial t} \vec{B}, \notag\\
    & \nabla\cdot \vec{B} = \nabla\cdot \vec{E} = 0.
\end{align}

Consider the simplest possible model of a medium: one that is dispersionless and lossless, with corresponding material properties specified by the positive constant dielectric permittivity $\varepsilon > 0$ and magnetic permeability $\mu > 0$, such that
\begin{align}\label{EqConstRel}
    \vec{D} = \varepsilon \vec{E}, \qquad  \vec{B} = \mu \vec{H}.
\end{align}

For such constitutive relations, Maxwell's Eq.~\eqref{EqMaxwellMacro} read
\begin{align}\label{EqMaxwellSimpleMacro}
    & \nabla\times \vec{H} =  \frac{\varepsilon}{c} \frac{\partial}{\partial t} \vec{E}, \notag\\
    & \nabla\times \vec{E} = -\frac{\mu}{c} \frac{\partial}{\partial t} \vec{H}, \notag\\
    & \nabla\cdot \vec{H} = \nabla\cdot \vec{E} = 0.
\end{align}

Note that macroscopic Eqs.~\eqref{EqMaxwellSimpleMacro} are obtained from the equations in vacuum~\eqref{EqMaxwellMicro} if the following substitution
\begin{align}\label{EqSubstitution}
    \begin{pmatrix}
        \vec{B}^{vac} \\
        \vec{E}^{vac} \\
        c
    \end{pmatrix}
    \longrightarrow
    \begin{pmatrix}
        \sqrt{\mu} \vec{H} \\
        \sqrt{\varepsilon} \vec{E} \\
        c / \sqrt{\varepsilon\mu}
    \end{pmatrix}.
\end{align}

Using this substitution, it is very easy to arrive at the generalization of Eq.~\eqref{EqSchwingerDefSpeedOfLight} to medium~\eqref{EqConstRel} can be readily obtained
\begin{align}\label{EqTowardNewDefSOL}
    \frac{d}{dt} \langle \vec{x} \rangle =  \frac{c^2}{\varepsilon\mu} \frac{\vec{P}}{E},
\end{align}
where
\begin{align}
    & \langle \vec{x} \rangle = \frac{1}{E} \int \vec{x} U d^3x,
    \qquad
    E = \int U d^3 x, \\
    & \vec{P} =  \varepsilon\mu \int \frac{\vec{S}}{c^2} d^3 x, \\
    & U = \left( \varepsilon\vec{E}^2 + \mu\vec{H}^2 \right) / (8\pi), \\
    & \vec{S} = c \vec{E} \times \vec{H} / (4\pi).
\end{align}
Equation~\eqref{EqTowardNewDefSOL} is the sought definition of the velocity of light in medium. Note that definition~\eqref{EqTowardNewDefSOL} originally appeared in Ref.~\cite{Real-27}, albeit using different notations. The principles and foundations of the derivation of this definition have been demonstrated in the monograph~\cite{Real-28}.

Furthermore, it follows from Eqs.~\eqref{EqMassVacuum} and \eqref{EqSubstitution} that the mass of light in medium
\begin{align}\label{EqMassInMedium}
    m = \frac{\mu\varepsilon}{c^2} E.
\end{align}

From Eq.~\eqref{EqTowardNewDefSOL}, we get
\begin{align}
    \left| \frac{d}{dt} \langle \vec{x} \rangle \right|
    &= c \left| \frac{\int \vec{E} \times \vec{H} d^3 x}{\frac{1}{2}\int \left( \varepsilon\vec{E}^2 + \mu\vec{H}^2 \right)d^3x  } \right| \label{EqExpandedVel} \\
    \leq &  c  \frac{\int \left| \vec{E} \times \vec{H}  \right| d^3 x}{\frac{1}{2}\int \left( \varepsilon\vec{E}^2 + \mu\vec{H}^2 \right)d^3x  }.
\end{align}
To further simply the last inequality, we can make use of the following
\begin{align}\label{EqVectInequality}
    |\vec{a} \times \vec{b}| \leq \frac{\vec{a}^2 +\vec{b}^2}{2}.
\end{align}
Inequality~\eqref{EqVectInequality} follows from the definition of the cross product $|\vec{a} \times \vec{b}| = |\vec{a} | |\vec{b}| \sin \theta \leq |\vec{a} | |\vec{b}|$ and the fact that the geometric mean of $|\vec{a} |^2$ and $|\vec{b} |^2$ cannot exceed their arithmetic mean.

Using inequality~\eqref{EqVectInequality}, we get
\begin{align}
    &\sqrt{\varepsilon\mu} | \vec{E} \times \vec{H}| \leq \frac{\varepsilon \vec{E}^2 +\mu \vec{b}^2}{2} \notag\\
    & \Longrightarrow \frac{\sqrt{\varepsilon\mu} \int \left| \vec{E} \times \vec{H}\right|d^3 x }{\frac{1}{2}\int \left(\varepsilon \vec{E}^2 +\mu \vec{H}^2\right)d^3 x} \leq 1.
\end{align}

Hence,
\begin{align}
     \left| \frac{d}{dt} \langle \vec{x} \rangle \right| \leq \frac{c}{\sqrt{\varepsilon\mu}}.
\end{align}
In other words, the speed of light definition~\eqref{EqTowardNewDefSOL} cannot exceed the phase velocity of light in medium.

Consider the special case of a linearly polarized plane wave
\begin{align}
    \vec{E} = E_0 \vec{i} f(kz - \omega t), \notag\\
    \vec{H} = H_0 \vec{j} f(kz - \omega t),
\end{align}
where $f(\cdot)$ is an arbitrary differential function, and $\vec{i}$ and $\vec{j}$ are units vectors along the $x$ and $y$ axis, respectively.

Substituting these expressions into Maxwell's equations~\eqref{EqMaxwellSimpleMacro}, we get
\begin{align}
    -k H_0 \vec{i}  f'(kz - \omega t) &= -\omega \frac{\varepsilon}{c}  E_0 \vec{i} f'(kz - \omega t), \notag\\
    k E_0 \vec{j} f'(kz - \omega t) &= \omega \frac{\mu}{c} H_0 \vec{j} f'(kz - \omega t) \Longrightarrow
\end{align}
\begin{align}
    H_0 = \omega \varepsilon E_0 / (ck),
    \qquad k E_0 = \mu\varepsilon E_0 \omega^2 / (c^2k)
\end{align}
Whence,
\begin{align}
    k= \sqrt{\mu\varepsilon}
    \omega / c,
    \qquad
    H_0 = E_0 \sqrt{\varepsilon/\mu}.
\end{align}
Substituting this solution into Eq.~\eqref{EqExpandedVel}, we get
\begin{align}\label{EqVelocityInMediumEqualsPhaseVel}
    \left| \frac{d}{dt} \langle \vec{x} \rangle \right|
    &= c \frac{E_0 H_0 \int  f^2(kz-\omega t) d^3 x}{\frac{1}{2}\left( \varepsilon E_0^2 + \mu H_0^2 \right) \int  f^2(kz-\omega t) d^3 x } \notag\\
    &= c \sqrt{\frac{\varepsilon}{\mu}} \frac{2 E_0^2}{\varepsilon E_0^2 + \varepsilon E_0^2} \notag\\
    &= \frac{c}{\sqrt{\mu\varepsilon}}
\end{align}
Therefore, for plane waves, the proposed velocity coincides with the phase velocity of light.

\section{Conclusion}

We have introduced formula~\eqref{EqTowardNewDefSOL} as a definition for the velocity of light in isotropic and homogeneous non-conducting media. This result interprets the velocity as the speed of energy transfer in the specified medium, presented in integral form as a functional of the electromagnetic fields. As a consequence, we have generalized the Einstein mass-energy relation to electromagnetic field in medium~\eqref{EqMassInMedium}.

According to the definition of mass~\eqref{EqMassInMedium}, electromagnetic field in materials with $\varepsilon\mu = 0$ is massless, irrespective of the field's intensity. Such materials, called near-zero-index materials~\cite{Real-29}, are a subject of active investigations in modern optics due to their unique properties -- one of which is that the phase velocity of light is infinite, hence the electrodynamics in such a medium reduces to electrostatics. Recently, it was experimentally shown~\cite{Real-30} that common liquids like water can be driven by an external laser field such that $\varepsilon = 0$ temporarily. The uncovered fact that light carries no mass and hence no momentum in such materials is another curious property of near-zero-index materials.

This work is presented in 2025, marking the centenary of the theory of quantum mechanics, see, e.g, ~\cite{Real-31}. After more than 120 years since the classical formulations, new perspectives on fundamental questions remain valuable. As noted in~\cite{Real-32}, modern considerations of light velocity face significant conceptual challenges beyond mere pedagogy. The derived formula~\eqref{EqTowardNewDefSOL} may find applications in contemporary studies of quantum fluids of light~\cite{Real-33} and dressed photons~\cite{Real-34}. Our approach demonstrates that Maxwell's equations and their well-established experimental basis provide a rich foundation for understanding electromagnetic phenomena, including new perspectives on the velocity and mass of light that complement the standard definitions.

\section*{Acknowledgment}

D.I.B. was supported by Army Research Office (ARO) (grant W911NF-23-1-0288; program manager Dr.~James Joseph). The views and conclusions contained in this document are those of the authors and should not be interpreted as representing the official policies, either expressed or implied, of ARO, or the U.S. Government. The U.S. Government is authorized to reproduce and distribute reprints for Government purposes notwithstanding any copyright notation herein.

\end{document}